\documentclass[conference]{IEEEtran}
\IEEEoverridecommandlockouts
\usepackage{cite}
\usepackage{amsmath,amssymb,amsfonts}
\usepackage{graphicx}
\usepackage{textcomp}
\usepackage{xcolor}
\usepackage{algorithm}
\usepackage{algorithmic}
\usepackage{booktabs}
\usepackage{multirow}
\usepackage{stfloats}
\usepackage{caption}
\usepackage{subcaption}
\def\BibTeX{{\rm B\kern-.05em{\sc i\kern-.025em b}\kern-.08em
    T\kern-.1667em\lower.7ex\hbox{E}\kern-.125emX}}

\begin{document}

\title{Diffusion-OAMP for Joint Image Compression and Wireless Transmission}

\author{\IEEEauthorblockN{Wentao Hou, Yimin Bai, Zelei Luo, Jiadong Hong, Lei Liu}
\IEEEauthorblockA{College of Information Science and Electronic Engineering, Zhejiang University, China \\
Email: \{wentao010, yimin.bai, zeleiluo, jiadong5, lei\_liu\}@zju.edu.cn}
}

\maketitle

\begin{abstract}
Joint image compression and wireless transmission remain relatively underexplored compared to
 generic image restoration, despite its importance in practical communication systems.
  We formulate this problem under an equivalent
linear model, and propose Diffusion-OAMP, a training-free reconstruction framework that embeds a pre-trained
diffusion model into the OAMP algorithm. In Diffusion-OAMP, the OAMP linear estimator produces pseudo-AWGN
observations, while the diffusion model serves as a nonlinear estimator under an SNR-matching
rule. This framework offers a way to incorporate multiple generative priors into OAMP. Experiments with varying compression ratios and noise levels show 
that Diffusion-OAMP performs favorably against classic methods in the evaluated settings.
\end{abstract}


\section{Introduction}

This paper considers the problem of joint image compression and wireless transmission,
which is relatively underexplored compared to generic image restoration, 
despite its relevance to practical and emerging intelligent communication systems~\cite{li2025large}.
The joint process can be described by the equivalent linear model:
\begin{equation}
\mathbf{y} = \mathbf{A}\mathbf{x} + \mathbf{n},
\label{eq:sys_model}
\end{equation}
where $\mathbf{y}\in\mathbb{R}^{M}$ denotes the received signal vector,
$\mathbf{x}\in\mathbb{R}^{N}$ denotes the transmitted signal,
$\mathbf{A}\in\mathbb{R}^{M\times N}$ is the equivalent matrix representing the joint effect of
modulation, compression, and transmission, and
$\mathbf{n}\sim\mathcal{N}(\mathbf{0},\sigma^2\mathbf{I})$ is additive white Gaussian noise
(AWGN). The recovery performance is therefore determined not only by the conditioning of
$\mathbf{H}$ and the noise level of $\mathbf{n}$, but also by how accurately the prior underlying image can be
incorporated into the receiver.

Approximate message passing (AMP) algorithms\cite{AMP1,AMP2} provide an attractive route to 
solve the linear recovery problem in \eqref{eq:sys_model}. In particular, Orthogonal AMP (OAMP)\cite{OAMP} and Memory AMP\cite{memoryamp} can achieve Replica-optimal performance. The linear / nonlinear
decoupling structure in the AMP-type algorithm is especially appealing for joint compression and transmission, and previous work has already shown that classical denoisers such as BM3D\cite{BM3D} can be incorporated into this framework\cite{turbocs}. However, the optimality of the AMP-type algorithm relies on having an accurate
prior model for $\mathbf{x}$ such as Bernoulli-Gaussian, while high-dimensional signals such as natural 
images do not admit a simple analytically tractable distribution. Once the assumed prior 
is mismatched to the true image distribution, the performance of OAMP becomes sub-optimal.

Recent progress in generative AI offers a practical way to address this prior-modeling
bottleneck. Diffusion models\cite{DDPM,DDIM,scorebasedgenerative,flowmatchinggenerativemodeling} can
capture rich image distributions far beyond the hand-crafted analytical prior distribution. This capability has motivated a series of diffusion-based solvers for linear inverse problems\cite{DDRM,DDNM,pnpflowplugandplayimagerestoration}. However, most of these diffusion-based methods are designed for generic image restoration rather than the joint image compression and wireless transmission scenario considered here. Among recent advances, Score-Based Turbo Message Passing (STMP)\cite{scorebasedturbomessagepassing} is particularly relevant, demonstrating the potential of the modern generative model for plug-and-play compressive image recovery. Different from STMP\cite{scorebasedturbomessagepassing}, we focus on joint image compression and wireless transmission with channel-induced ill-conditioning, and develop a diffusion-assisted OAMP receiver. Our framework can also be instantiated with different diffusion models, such as DDIM\cite{DDIM} and Flow Matching\cite{flowmatchinggenerativemodeling}.

The main contributions of this paper are summarized as follows:
\begin{itemize}
    \item We propose Diffusion-OAMP, a training-free framework for joint image compression and wireless
    transmission that embeds a diffusion model into the OAMP algorithm.
    \item We show that the proposed framework is compatible with multiple diffusion priors. In this
    paper, we instantiate the nonlinear estimator with both DDIM and Flow Matching models and use
    an SNR-matching rule to select the diffusion time step at each OAMP iteration.
    \item Experiments on the CelebA dataset under multiple compression ratios and noise levels show
    that the proposed approach compares favorably with OAMP-BM3D and DDRM in the evaluated
    settings.
\end{itemize}

\section{Preliminaries}

\subsection{Linear System Model for Joint Image Compression and Wireless Transmission}

Modulation and compression of the original signal bits can be modeled as a linear
process from the source image vector $\mathbf{s}\in\mathbb{R}^{N}$ to the transmitted
signal:
\begin{equation}
\mathbf{x} = \mathbf{F}\mathbf{s},
\label{eq:rm_compression}
\end{equation}
where $\mathbf{F}$ is the joint modulation and compression operator. The source image
vector is assumed to be drawn from an image prior distribution $P_s$, denoted by
$\mathbf{s}\sim P_s$. The concrete construction of $\mathbf{F}$, based on
random multiplexing (RM) \cite{randommultiplexing}, is described in
Section~\ref{subsec:rm}.

Image transmission over the wireless channel can also be represented as a linear process:
\begin{equation}
\mathbf{y} = \mathbf{A}\mathbf{x} + \mathbf{n},
\label{eq:channel_model}
\end{equation}
where $\mathbf{y}\in\mathbb{R}^{M}$ denotes the received signal vector, $\mathbf{A}$ is the
channel matrix, and $\mathbf{n}\sim\mathcal{N}(\mathbf{0},\sigma^2\mathbf{I})$ is AWGN.

This paper views joint image compression and wireless transmission as an integrated
linear inverse problem by combining the two linear stages:
\begin{equation}
 \mathbf{y} = \mathbf{A}\underbrace{(\mathbf{F}\mathbf{s})}_{\mathbf{x}} + \mathbf{n}
 = \mathbf{A}\mathbf{x} + \mathbf{n},
\label{eq:linear_model}
\end{equation}
or equivalently $\mathbf{y}=(\mathbf{A}\mathbf{F})\mathbf{s}+\mathbf{n}$. From this point of view, the receiver recovers the source image $\mathbf{s}$ from $\mathbf{y}$ under the operator $\mathbf{A}\mathbf{F}$, while still exploiting the explicit intermediate variable $\mathbf{x}$ to separate channel estimation in the transmission domain from the image reconstruction in the source domain.

\subsection{AMP Algorithms for Linear System}

AMP algorithms~\cite{AMP1,AMP2} are a class of effective iterative methods for linear inverse problems, such as compressed sensing. By passing messages between a linear measurement module and a nonlinear estimation module, 
AMP-type algorithms turn a large coupled recovery problem into a sequence of simpler denoising and linear estimation steps.

OAMP\cite{OAMP} and MAMP\cite{memoryamp} further improve this framework by introducing an orthogonalization mechanism between the linear and nonlinear estimators. Compared with conventional AMP, OAMP is better suited to sensing matrices beyond the i.i.d. Gaussian case, admits rigorous state-evolution analyses, and can achieve Replica-optimal performance in the large-system limit for right-unitarily invariant matrices.

In this work, OAMP serves as the backbone algorithm for solving the integrated joint compression-and-transmission inverse problem in Eq.~\eqref{eq:linear_model}. The OAMP algorithm is performed in the transmission domain $\mathbf{x}$: its Linear Minimal Mean Square Error (LMMSE) estimator enforces consistency with the known channel model and the noisy observation, while the orthogonalized output is subsequently mapped to the source-image domain for diffusion-based nonlinear estimation.

Specifically, given the pseudo-Gaussian prior estimate
$(\mathbf{x}_{pri}^{\text{mmse}}, v_{pri}^{\text{mmse}})$, the OAMP LMMSE
performs an update under $\mathbf{y}=\mathbf{A}\mathbf{x}+\mathbf{n}$. The posterior
mean and covariance are
\begin{equation}
    \mathbf{x}_{post} = \mathbf{x}_{pri}^{\text{mmse}} +
    v_{pri}^{\text{mmse}}\mathbf{A}^{T}\mathbf{C}
    (\mathbf{y} - \mathbf{A}\mathbf{x}_{pri}^{\text{mmse}}),
\end{equation}
where $\mathbf{C}\triangleq
\left(\sigma^{2}\mathbf{I}+v_{pri}^{\text{mmse}}\mathbf{A}\mathbf{A}^{T}\right)^{-1}$.
\begin{equation}
    \mathbf{V}_{post} = v_{pri}^{\text{mmse}}\mathbf{I} -
    (v_{pri}^{\text{mmse}})^{2}\mathbf{A}^{T}
    \left(\sigma^2\mathbf{I}+v_{pri}^{\text{mmse}}\mathbf{A}\mathbf{A}^T\right)^{-1}\mathbf{A},
    \label{eq:le_variance}
\end{equation}
with scalar variance
\begin{equation}
v_{post} \triangleq \frac{1}{M}\mathrm{tr}(\mathbf{V}_{post}).
\label{eq:scalar_var}
\end{equation}

OAMP reduces the correlation issue between the Linear Estimator (LE) and the Non-Linear Estimator (NLE) module through an
orthogonalization step, which converts
$(\mathbf{x}_{post}, v_{post})$ into
$(\mathbf{x}_{post}^{\text{orth}}, v_{post}^{\text{orth}})$:
\begin{equation}
 v_{post}^{\text{orth}} =
 \left(v_{post}^{-1} - (v_{pri}^{\text{mmse}})^{-1}\right)^{-1},
 \label{eq:variance_orth}
\end{equation}
\begin{equation}
\mathbf{x}_{post}^{\text{orth}} =
v_{post}^{\text{orth}}
\left(
v_{post}^{-1}\mathbf{x}_{post} -
(v_{pri}^{\text{mmse}})^{-1}\mathbf{x}_{pri}^{\text{mmse}}
\right).
\label{eq:mean_orth}
\end{equation}
The pair $(\mathbf{x}_{post}^{\text{orth}}, v_{post}^{\text{orth}})$ serves as an extrinsic
pseudo-observation that connects channel-aware estimation in the $\mathbf{x}$ domain with
generative reconstruction in the $\mathbf{s}$ domain.

\subsection{Random Multiplexing}
\label{subsec:rm}

In image compressed sensing, a classical design is the partial DFT sensing matrix $\boldsymbol{\Phi}_{\mathrm{pDFT}}=\mathbf{S}\mathbf{F}_{\mathrm{DFT}}$ used in Turbo compressed sensing~\cite{turbocs}, where $\mathbf{S}$ randomly selects and reorders rows of a unitary DFT matrix. Random Multiplexing (RM)~\cite{randommultiplexing}, developed for wireless communication, instead applies a channel-independent random transform before transmission. This transform makes the equivalent channel more suitable for the AMP-type receivers.

Inspired by this idea, we use one operator to perform both compression and multiplexing:
\begin{equation}
    \mathbf{F} = \mathbf{S}\boldsymbol{\Xi},
    \qquad
    \boldsymbol{\Xi}=\mathbf{P}\mathbf{T}_{\text{DCT}}\mathbf{D},
    \label{eq:rm_matrix}
\end{equation}
where $\mathbf{D}$ is a diagonal random sign matrix, $\mathbf{T}_{\text{DCT}}$ is an
orthogonal DCT matrix, $\mathbf{P}$ is a random permutation matrix, and
$\mathbf{S}\in\mathbb{R}^{M\times N}$ selects $M$ rows, with compression ratio
$\beta=M/N$. Thus,  $\boldsymbol{\Xi}$ provides the RM-style multiplexing, while
$\mathbf{S}$ provides compression, yielding the transmitted vector
$\mathbf{x}=\mathbf{F}\mathbf{s}$. Since $\mathbf{F}$ is rectangular,
$\mathbf{F}^{-1}$ in the receiver denotes the corresponding zero-filled inverse
$\boldsymbol{\Xi}^{T}\mathbf{S}^{T}$.

\subsection{Diffusion Models}

Diffusion~\cite{DDPM} and Flow Matching~\cite{flowmatchinggenerativemodeling} models have emerged as a powerful class of generative AI, achieving state-of-the-art performance in high-quality image synthesis. Unlike traditional hand-crafted priors, diffusion models implicitly capture the underlying data distribution through an iterative denoising process. This intrinsic ability to learn complex data structures makes them an ideal mechanism for providing high-fidelity empirical priors in solving inverse problems.

Beyond pure image generation, diffusion models have also been directly applied to solve linear inverse problems. Representative examples include DDRM~\cite{DDRM}, DDNM~\cite{DDNM}, and PnP-Flow~\cite{pnpflowplugandplayimagerestoration}. However, these methods, as well as the closely related STMP~\cite{scorebasedturbomessagepassing}, focus mainly on conventional image restoration or compressive image recovery. They are not specifically designed for joint image compression and wireless transmission. Motivated by this gap, we embed diffusion priors into the OAMP receiver for joint image compression and wireless transmission. The resulting framework uses the diffusion model as a plug-in nonlinear estimator and supports different priors, including DDIM and Flow Matching.

\begin{algorithm}[t]
\caption{RM Transmission and Diffusion-OAMP Reception}
\label{alg:Diffusion-OAMP}
\begin{algorithmic}[1]
\renewcommand{\algorithmicrequire}{\textbf{Input:}}
\renewcommand{\algorithmicensure}{\textbf{Output:}}
\REQUIRE Source image signal $\mathbf{s}$, joint RM-compression operator $\mathbf{F}$, channel $\mathbf{A}$, noise variance $\sigma^2$, DM denoiser $\phi(\cdot,t)$, maximum iteration number $K$, tolerance $\tau$.
\STATE \textbf{Joint RM and compression:} $\mathbf{F}\mathbf{s} \rightarrow \mathbf{x}$
\STATE \textbf{Transmission:} $\mathbf{A}\mathbf{x} + \mathbf{n} \rightarrow \mathbf{y}$
\STATE \textbf{Initialization:} $\left(\mathbf{0}, \frac{1}{M}\|\mathbf{y}\|_2^2\right) \rightarrow (\mathbf{x}_{pri}^{\text{mmse}}, v_{pri}^{\text{mmse}})$
\FOR{$k = 1, \dots, K$}
    \STATE \textbf{Linear Estimator:} $(\mathbf{x}_{pri}^{\text{mmse}}, v_{pri}^{\text{mmse}}) \rightarrow (\mathbf{x}_{post}, v_{post})$
    \STATE \textbf{Orthogonalization:} $(\mathbf{x}_{post}, v_{post}) \rightarrow (\mathbf{x}_{post}^{\text{orth}}, v_{post}^{\text{orth}})$
    \STATE \textbf{Inverse RM:} $\mathbf{F}^{-1}\mathbf{x}_{post}^{\text{orth}} \rightarrow \mathbf{s}_{in}$
    \STATE \textbf{SNR Matching:} $v_{post}^{\text{orth}} \rightarrow t^*$
    \STATE \textbf{NLE:} $(\mathbf{s}_{in}, t^*) \rightarrow \phi_\perp(\mathbf{s}_{in})$
    \STATE \textbf{Forward RM:} $\mathbf{F}\phi_\perp(\mathbf{s}_{in}) \rightarrow \tilde{\mathbf{x}}$
    \STATE \textbf{MMSE Correction:} $(\tilde{\mathbf{x}}, \mathbf{x}_{post}^{\text{orth}}) \rightarrow (\mathbf{x}_{pri}^{\text{mmse}}, v_{pri}^{\text{mmse}})$
    \IF{relative change of $\mathbf{x}_{pri}^{\text{mmse}} < \tau$}
        \STATE \textbf{break}
    \ENDIF
\ENDFOR
\STATE \textbf{Inverse RM:} $\mathbf{F}^{-1}\mathbf{x}_{pri}^{\text{mmse}} \rightarrow \hat{\mathbf{s}}$
\ENSURE Reconstructed source signal $\hat{\mathbf{s}}$
\end{algorithmic}
\end{algorithm}

\begin{figure*}[t]
    \centering
    \includegraphics[width=\textwidth]{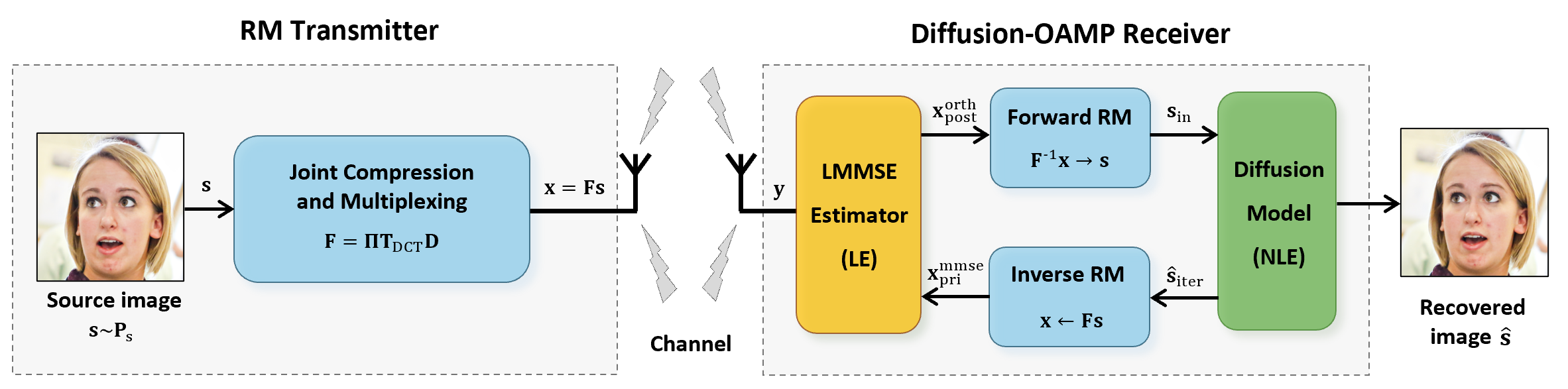}
    \caption{Overall system diagram of the proposed framework. Left: the source image
    $\mathbf{s}$ is mapped by the joint random compression and multiplexing operator to
    produce $\mathbf{x}$, then transmitted
    through a wireless channel corrupted by AWGN. Right: the Diffusion-OAMP receiver alternates among an
    LMMSE estimator, inverse/forward RM transforms between $\mathbf{x}$ and $\mathbf{s}$, and a
    diffusion-model prior to reconstruct the image.}
    \label{fig:framework}
\end{figure*}

\section{Diffusion-OAMP Algorithm}

\subsection{Algorithm Overview}

The overall system consists of an RM transmitter, a wireless channel, and a Diffusion-OAMP receiver. The RM transmitter produces $\mathbf{x}=\mathbf{F}\mathbf{s}$, and the wireless channel generates $\mathbf{y}=\mathbf{A}\mathbf{x}+\mathbf{n}$. The receiver iterates the OAMP linear estimator, inverse RM, SNR matching, diffusion-based nonlinear estimation, forward RM, and MMSE correction until convergence. The complete procedure is summarized in
Algorithm~\ref{alg:Diffusion-OAMP}.

\subsection{Diffusion NLE}
We instantiate the generative prior using two distinct diffusion-based methods, namely DDIM~\cite{DDIM} and Flow Matching~\cite{flowmatchinggenerativemodeling}.

\textbf{DDIM}~\cite{DDIM} defines a non-Markovian forward process that maps a clean image
$\mathbf{s}_0$ to a noisy sample:
\begin{equation}
    \mathbf{s}_t = \sqrt{\bar{\alpha}_t}\,\mathbf{s}_0 +
    \sqrt{1-\bar{\alpha}_t}\,\boldsymbol{\epsilon},
    \qquad \boldsymbol{\epsilon} \sim \mathcal{N}(\mathbf{0},\mathbf{I}),
    \label{eq:DDIMforward}
\end{equation}
where $\bar{\alpha}_t$ is a decreasing noise schedule. A neural network
$\boldsymbol{\epsilon}_\theta$ is trained to predict the added noise. Given $\mathbf{s}_t$, a deterministic DDIM reverse update can be written as
\begin{equation}
\hat{\mathbf{s}}_0 =
\frac{\mathbf{s}_t - \sqrt{1-\bar{\alpha}_t}\,\boldsymbol{\epsilon}_\theta(\mathbf{s}_t,t)}
{\sqrt{\bar{\alpha}_t}},
\label{eq:ddim_x0}
\end{equation}
\begin{equation}
\mathbf{s}_{t-1} =
\sqrt{\bar{\alpha}_{t-1}}\,\hat{\mathbf{s}}_0 +
\sqrt{1-\bar{\alpha}_{t-1}}\,\boldsymbol{\epsilon}_\theta(\mathbf{s}_t,t).
\label{eq:ddim_reverse}
\end{equation}

\textbf{Flow Matching (FM)}~\cite{flowmatchinggenerativemodeling} offers a simpler
probability path from noise to data:
\begin{equation}
    \mathbf{z}_t = (1-t)\,\boldsymbol{\epsilon} + t\,\mathbf{s}_0, \qquad t \in [0,1],
    \label{eq:FMforward}
\end{equation}
where a network $\mathbf{v}_\theta$ is trained to regress the transport velocity from the path
state $\mathbf{z}_t$ and time $t$. Signal recovery is then written as the ODE
\begin{equation}
    \frac{d\mathbf{z}_t}{dt} = \mathbf{v}_\theta(\mathbf{z}_t, t),
    \quad \mathbf{z}_0 = \boldsymbol{\epsilon}, \quad \mathbf{z}_1 = \mathbf{s}_0.
    \label{eq:fm_ode}
\end{equation}

\subsection{Inverse RM and SNR Matching}

After the OAMP LE and orthogonalization steps, the orthogonalized estimate is mapped back to the
source-image domain:
\begin{equation}
\mathbf{s}_{in} = \mathbf{F}^{-1}\mathbf{x}_{post}^{\text{orth}}.
\label{eq:inverse_rm}
\end{equation}
Since the multiplexing component of $\mathbf{F}$ is orthogonal, the AWGN structure is preserved
under the inverse RM transform. Hence,
\begin{equation}
\mathbf{s}_{in} = \mathbf{s} + \sqrt{v_{post}^{\text{orth}}}\,\boldsymbol{\epsilon},
\qquad
\boldsymbol{\epsilon}\sim\mathcal{N}(\mathbf{0},\mathbf{I}),
\label{eq:s_in_awgn}
\end{equation}
which matches the forward corruption model used by the diffusion prior.

We select the diffusion time step by matching the effective OAMP error variance to the diffusion
noise level:
\begin{equation}
t^*=
\begin{cases}
\dfrac{1}{1 + \sqrt{v_{post}^{\text{orth}}}} & \text{(Flow Matching)},\\[6pt]
\dfrac{1}{1 + v_{post}^{\text{orth}}} & \text{(DDIM)}.
\end{cases}
\label{eq:snr_matching}
\end{equation}
As $v_{post}^{\text{orth}}$ decreases across iterations, $t^*$ moves toward the clean end of the
diffusion trajectory, so the NLE is called on progressively less corrupted observations.

\subsection{MMSE Correction}

Given $t^*$, the NLE evaluates the pre-trained denoiser $\phi(\mathbf{s}_{in}, t^*)$ and projects the estimation onto the learned image manifold. To produce an orthogonalized message for the next OAMP iteration, we use Stein's Unbiased Risk Estimator (SURE) to construct
\begin{equation}
    \phi_\perp(\mathbf{s}_{in}) =
    \phi(\mathbf{s}_{in},t^*) - \phi'(\mathbf{s}_{in})\,\mathbf{s}_{in},
    \label{eq:mmse_direction}
\end{equation}
where the scalar divergence estimate $\phi'(\mathbf{s}_{in})$ is computed by a single Monte Carlo
perturbation with $\mathbf{w}\sim\mathcal{N}(\mathbf{0},\mathbf{I})$:
\begin{equation}
    \phi'(\mathbf{s}_{in})=
    \lim_{\epsilon\to0}
    \frac{\langle \mathbf{w},
    \phi(\mathbf{s}_{in}+\epsilon\mathbf{w},t^*)-
    \phi(\mathbf{s}_{in},t^*)\rangle}{\epsilon}.
\end{equation}

The denoised image estimate is then remodulated,
\begin{equation}
\tilde{\mathbf{x}} = \mathbf{F}\phi_\perp(\mathbf{s}_{in}),
\label{eq:forward_rm}
\end{equation}
and a posterior MMSE correction maps $\tilde{\mathbf{x}}$ back to the next prior pair
$(\mathbf{x}_{pri}^{\text{mmse}}, v_{pri}^{\text{mmse}})$:
\begin{equation}
\mathbf{x}_{pri}^{\text{mmse}} = \beta^* \tilde{\mathbf{x}}, \quad
v_{pri}^{\text{mmse}} = \frac{1}{M}\|\mathbf{A}\mathbf{x}_{pri}^{\text{mmse}} - \mathbf{y}\|_2^2.
\label{eq:mmse_correction}
\end{equation}
Where $\beta^* = \frac{\langle \tilde{\mathbf{x}}, \mathbf{x}_{post}^{\text{orth}} \rangle}
{\|\tilde{\mathbf{x}}\|_2^2}$.

\begin{figure}[t]
    \centering
    \includegraphics[width=\linewidth]{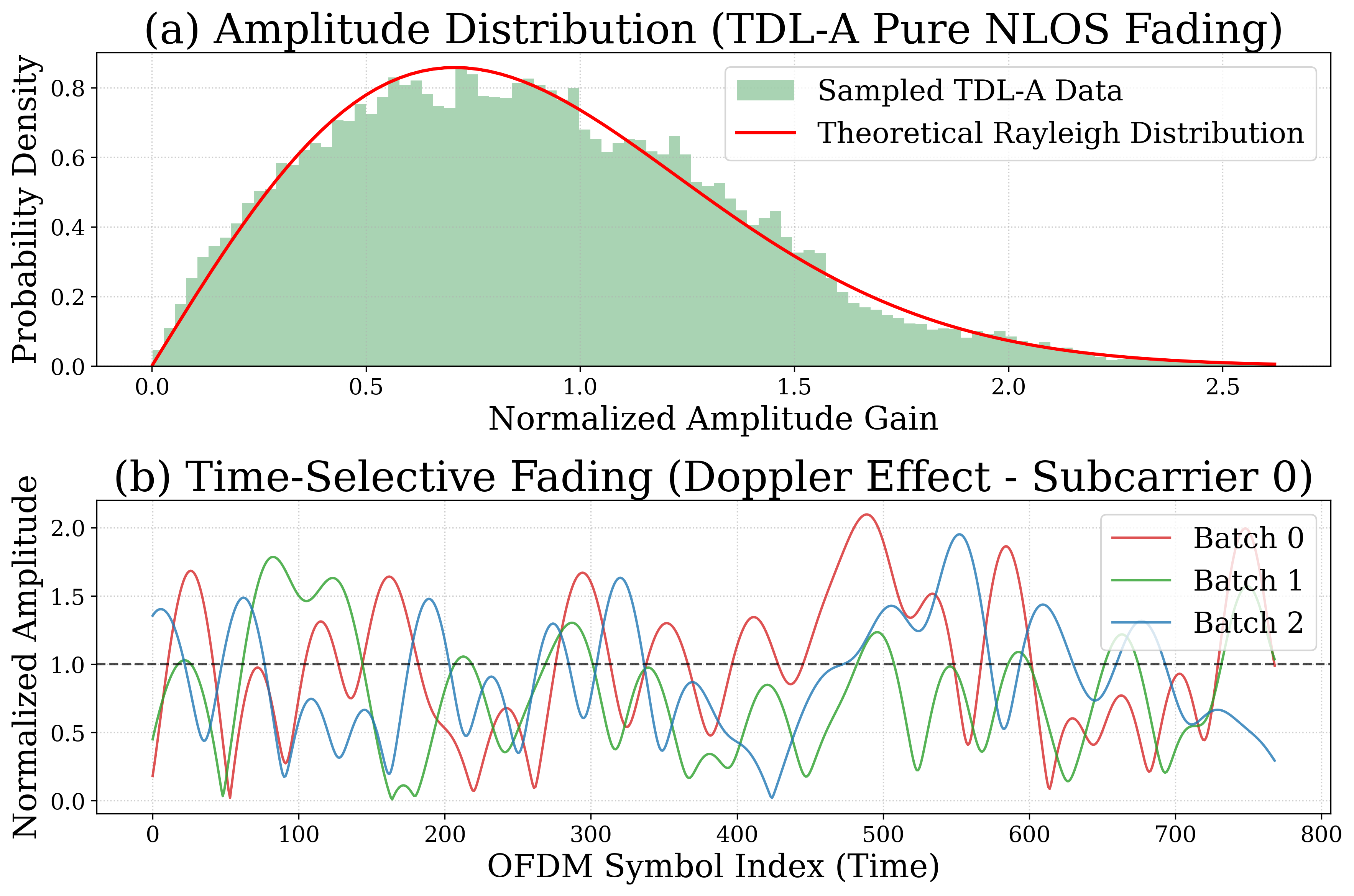} 
    \caption{Visualization of the simulated 3GPP TR 38.901 TDL-A channel characteristics. (a) The empirical amplitude distribution of the generated pure NLOS fading channel tightly matches the theoretical Rayleigh distribution. (b) Time-selective fading variations induced by the Doppler effect across $769$ consecutive OFDM symbols. The underlying OFDM system is configured with a $3.5$ GHz carrier frequency, $30$ kHz subcarrier spacing, a $256$-point FFT, a cyclic prefix length of $18$, and a UE mobility of $30$ km/h.}
    \label{fig:channel_model}
\end{figure}

\section{Experiments}
This section evaluates the proposed Diffusion-OAMP framework across four dimensions: 1) quantitative recovery in pure image compression under ideal AWGN conditions; 2) quantitative recovery in joint compression and transmission over 3GPP TR 38.901~\cite{3GPP} Tapped Delay Line (TDL) fading channels; 3) visual assessments of perceptual fidelity and structural preservation; and 4) convergence performance across successive OAMP iterations. By systematically comparing our approach with baseline methods under diverse compression ratios and noise levels, we validate the framework's effectiveness in mitigating complex channel distortions.

\begin{figure*}[t]
    \centering
    \includegraphics[width=\textwidth]{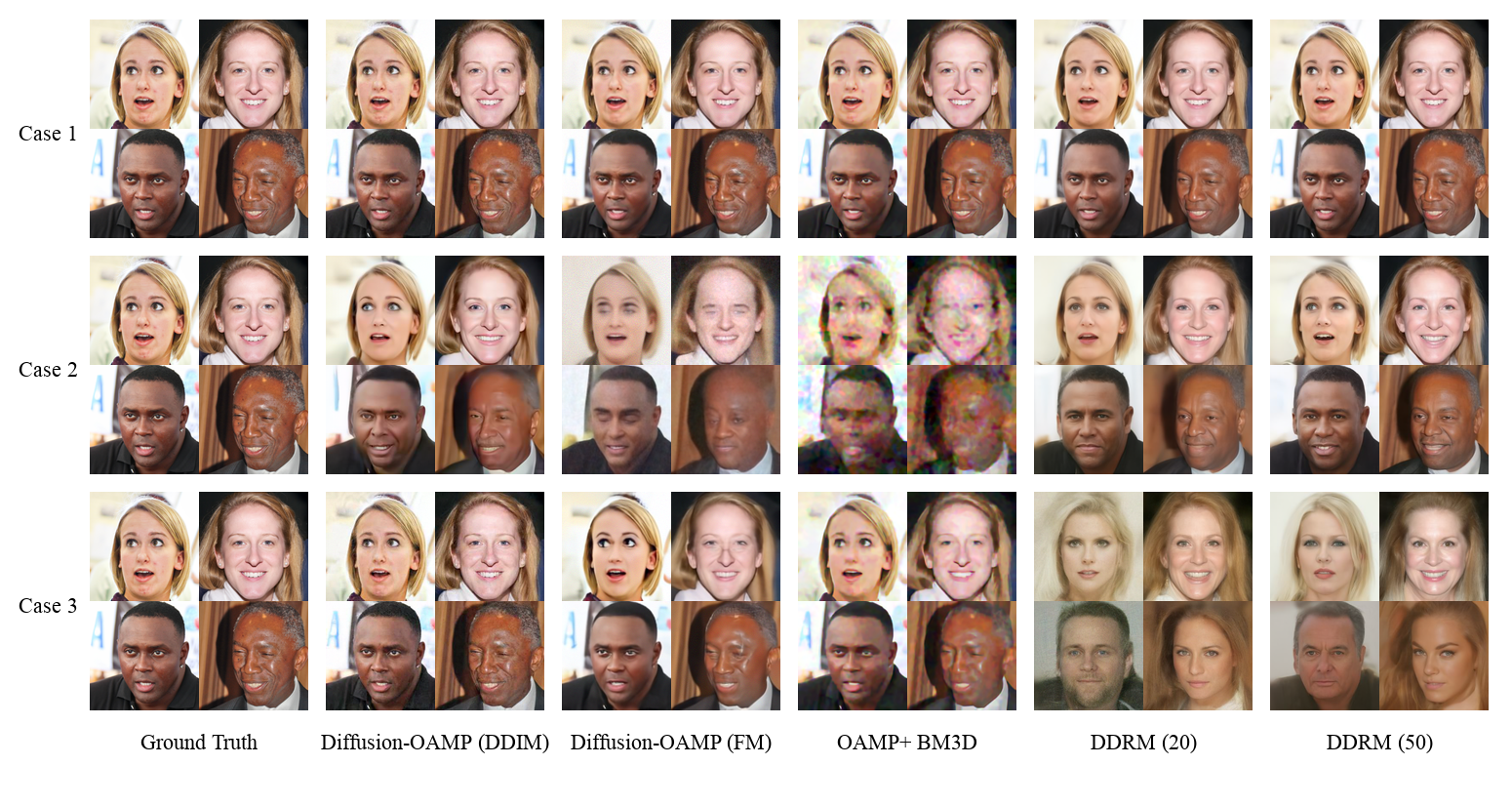} 
    \caption{Visual comparison on CelebA over physical TDL-A fading channels under three representative conditions. Case 1 (mild): $\beta=0.7, \sigma=0.05$; Case 2 (high-noise): $\beta=0.4, \sigma=0.5$; and Case 3 (low-compression): $\beta=0.1, \sigma=0.05$. Notably, our proposed Diffusion-OAMP(DDIM) framework consistently preserves high-frequency facial details and identity characteristics, avoiding the severe artifacts or structural distortions evident in baseline methods under harsh regimes.}
    \label{fig:visual}
\end{figure*}

Fig.~\ref{fig:visual} provides detailed visual comparisons under various degradation cases, including mild, high-noise, and low-compression settings. These visual results complement the quantitative evaluation, further highlighting the inherent robustness of our Diffusion-OAMP framework. Compared with conventional OAMP+BM3D and DDRM, our variants better preserve facial structures and fine-grained textures. By effectively suppressing over-smoothing, unwanted color shifts, and structural artifacts, the results clearly demonstrate the effectiveness of the proposed model-based diffusion reconstruction scheme.

\subsection{Experimental Setup}
We utilize the CelebA~\cite{liu2015faceattributes} dataset ($256\times256$) with a compression ratio defined as $\beta=M/N$. The equivalent observation matrix is assumed to be right-unitarily invariant, and AWGN levels are set to $\sigma \in \{0.05, 0.5\}$. For the joint compression and transmission scenario, realistic TDL-A fading channels are constructed via the Sionna~\cite{hoydis2022sionna} library, with specific OFDM configurations and visual channel characteristics detailed in Fig.~\ref{fig:channel_model}.

\subsection{Quantitative Performance in Pure Image Compression}
Table~\ref{tab:quantitative_awgn} summarizes the recovery metrics under ideal AWGN conditions. The proposed Diffusion-OAMP framework, instantiated with DDIM and Flow Matching priors, consistently outperforms traditional baselines such as DDRM~\cite{DDRM} and OAMP+BM3D~\cite{turbocs} across all evaluated compression ratios ($\beta$) and noise levels ($\sigma$). Specifically, Diffusion-OAMP (DDIM) achieves the highest PSNR, particularly in high-noise regimes ($\sigma = 0.5$).

% --- Table 1: Pure Image Compression ---
\begin{table}[t]
\centering
\caption{Quantitative results for pure image compression under different compression ratios and AWGN levels.}
\renewcommand{\arraystretch}{1.2}
\setlength{\tabcolsep}{4pt}
\footnotesize
\resizebox{\columnwidth}{!}{%
\begin{tabular}{clcccc}
\toprule
\multirow{2}{*}{$\beta$} & \multirow{2}{*}{Method} & \multicolumn{2}{c}{$\sigma = 0.05$} & \multicolumn{2}{c}{$\sigma = 0.5$} \\
\cmidrule(lr){3-4} \cmidrule(lr){5-6}
& & PSNR $\uparrow$ & SSIM $\uparrow$ & PSNR $\uparrow$ & SSIM $\uparrow$ \\
\midrule
\multirow{5}{*}{0.7}
& \mbox{Diffusion-OAMP (DDIM)~(ours)} & 33.34 & 0.8930 & \textbf{25.99} & \textbf{0.7645} \\
& \mbox{Diffusion-OAMP (FM)~(ours)} & 33.18 & \textbf{0.9326} & 23.92 & 0.7066 \\
& \mbox{OAMP+BM3D~\cite{turbocs}} & \textbf{33.77} & 0.9160 & 23.05 & 0.5996 \\
& \mbox{DDRM~(20)~\cite{DDRM}} & 31.97 & 0.8978 & 24.50 & 0.7025 \\
& \mbox{DDRM~(50)~\cite{DDRM}} & 32.88 & 0.9107 & 24.85 & 0.7131 \\
\midrule
\multirow{5}{*}{0.4}
& \mbox{Diffusion-OAMP (DDIM)~(ours)} & \textbf{32.62} & 0.8952 & \textbf{24.66} & \textbf{0.7235} \\
& \mbox{Diffusion-OAMP (FM)~(ours)} & 31.69 & \textbf{0.9125} & 22.83 & 0.6501 \\
& \mbox{OAMP+BM3D~\cite{turbocs}} & 32.05 & 0.8854 & 21.66 & 0.5438 \\
& \mbox{DDRM~(20)~\cite{DDRM}} & 28.16 & 0.8259 & 22.32 & 0.6321 \\
& \mbox{DDRM~(50)~\cite{DDRM}} & 30.17 & 0.8629 & 23.24 & 0.6575 \\
\midrule
\multirow{5}{*}{0.1}
& \mbox{Diffusion-OAMP (DDIM)~(ours)} & \textbf{30.30} & \textbf{0.8640} & \textbf{21.69} & \textbf{0.6258} \\
& \mbox{Diffusion-OAMP (FM)~(ours)} & 28.38 & 0.8445 & 17.70 & 0.4663 \\
& \mbox{OAMP+BM3D~\cite{turbocs}} & 28.08 & 0.7856 & 17.58 & 0.3783 \\
& \mbox{DDRM~(20)~\cite{DDRM}} & 17.35 & 0.5045 & 15.48 & 0.4367 \\
& \mbox{DDRM~(50)~\cite{DDRM}} & 21.02 & 0.6259 & 18.00 & 0.5105 \\
\bottomrule
\end{tabular}
}
\label{tab:quantitative_awgn}
\end{table}

\begin{figure}[t]
    \centering
    \includegraphics[width=\linewidth]{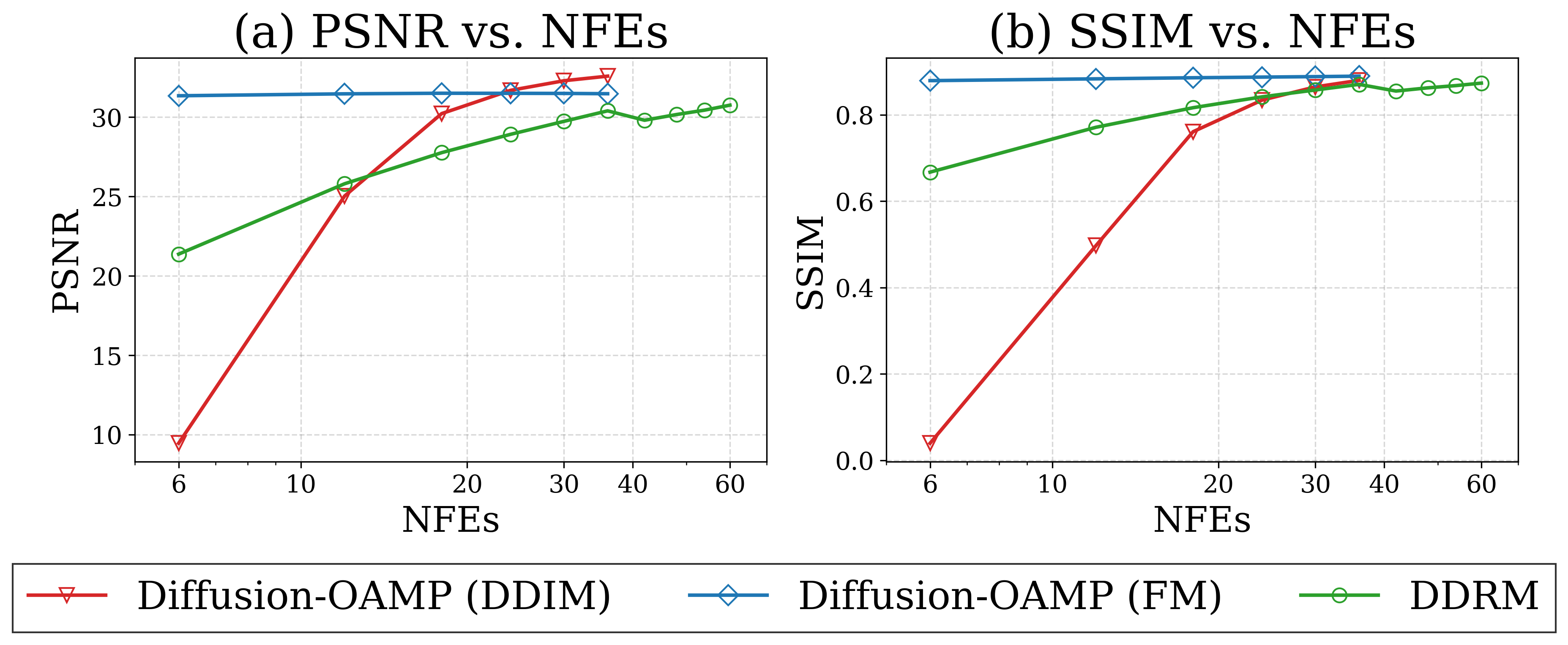} 
    \caption{Performance trade-off regarding the Number of Function Evaluations (NFEs). (a) PSNR vs. NFEs and (b) SSIM vs. NFEs. Both Diffusion-OAMP (FM) and Diffusion-OAMP (DDIM) exhibit distinct convergence behaviors and enhanced sampling efficiency compared to DDRM.}
    \label{fig:nfe}
\end{figure}

To evaluate sampling efficiency, we further analyze the trade-off between reconstruction quality and the Number of Function Evaluations (NFEs). As illustrated in Fig.~\ref{fig:nfe}, Diffusion-OAMP (FM) exhibits rapid convergence, reaching near-optimal PSNR and SSIM within only 6 NFEs. While Diffusion-OAMP (DDIM) requires a higher NFE count to reach its performance plateau, it ultimately yields a higher peak PSNR. Both variants demonstrate enhanced sampling efficiency compared to DDRM, necessitating fewer neural network evaluations to achieve competitive metrics.
\subsection{Quantitative Performance in Joint Image Compression and Wireless Transmission}
Table ~\ref{tab:quantitative_fading} presents the performance averaged over 10 TDL-A fading channel realizations. In this scenario, time-selective fading renders the equivalent observation matrix severely ill-conditioned, leading to noticeable performance degradation for conventional solvers.

In contrast, the Diffusion-OAMP framework maintains stable reconstruction under these complex channel distortions. Specifically, Diffusion-OAMP (DDIM) yields the highest PSNR across most configurations, showing particular effectiveness in the challenging low-compression ($\beta=0.1$) and high-noise ($\sigma=0.5$) case. Diffusion-OAMP (FM) retains competitive structural preservation (SSIM) under milder conditions. These results indicate that embedding generative priors as non-linear estimators within OAMP effectively regularizes the ill-posed inversion problem induced by fading channels.

% --- Table 2: Joint Image Compression and Wireless Transmission ---
\begin{table}[t]
\centering
\caption{Quantitative results for joint image compression and wireless transmission (averaged over 10 selected TDL-A channels).}
\renewcommand{\arraystretch}{1.2}
\setlength{\tabcolsep}{4pt}
\footnotesize
\resizebox{\columnwidth}{!}{%
\begin{tabular}{clcccc}
\toprule
\multirow{2}{*}{$\beta$} & \multirow{2}{*}{Method} & \multicolumn{2}{c}{$\sigma = 0.05$} & \multicolumn{2}{c}{$\sigma = 0.5$} \\
\cmidrule(lr){3-4} \cmidrule(lr){5-6}
& & PSNR $\uparrow$ & SSIM $\uparrow$ & PSNR $\uparrow$ & SSIM $\uparrow$ \\
\midrule
\multirow{5}{*}{0.7}
& \mbox{Diffusion-OAMP (DDIM)~(ours)} & \textbf{33.15} & 0.8855 & \textbf{25.86} & \textbf{0.7715} \\
& \mbox{Diffusion-OAMP (FM)~(ours)} & 32.85 & \textbf{0.9206} & 22.04 & 0.4861 \\
& \mbox{OAMP+BM3D~\cite{turbocs}} & 31.32 & 0.9106 & 20.88 & 0.5468 \\
& \mbox{DDRM~(20)~\cite{DDRM}} & 31.19 & 0.8845 & 24.10 & 0.6886 \\
& \mbox{DDRM~(50)~\cite{DDRM}} & 32.41 & 0.9038 & 24.48 & 0.7006 \\
\midrule
\multirow{5}{*}{0.4}
& \mbox{Diffusion-OAMP (DDIM)~(ours)} & \textbf{32.25} & 0.8725 & \textbf{24.74} & \textbf{0.7439} \\
& \mbox{Diffusion-OAMP (FM)~(ours)} & 30.86 & 0.8691 & 19.98 & 0.3412 \\
& \mbox{OAMP+BM3D~\cite{turbocs}} & 29.56 & \textbf{0.8795} & 19.60 & 0.4850 \\
& \mbox{DDRM~(20)~\cite{DDRM}} & 27.60 & 0.8105 & 21.95 & 0.6233 \\
& \mbox{DDRM~(50)~\cite{DDRM}} & 29.74 & 0.8532 & 22.75 & 0.6415 \\
\midrule
\multirow{5}{*}{0.1}
& \mbox{Diffusion-OAMP (DDIM)~(ours)} & \textbf{29.63} & \textbf{0.8115} & \textbf{21.95} & \textbf{0.6484} \\
& \mbox{Diffusion-OAMP (FM)~(ours)} & 26.66 & 0.7483 & 17.39 & 0.2133 \\
& \mbox{OAMP+BM3D~\cite{turbocs}} & 24.83 & 0.7295 & 16.20 & 0.3392 \\
& \mbox{DDRM~(20)~\cite{DDRM}} & 16.51 & 0.5254 & 14.73 & 0.4349 \\
& \mbox{DDRM~(50)~\cite{DDRM}} & 21.05 & 0.6605 & 17.40 & 0.5263 \\
\bottomrule
\end{tabular}
}
\label{tab:quantitative_fading}
\end{table}

\subsection{Convergence Performance}
\begin{figure}[t]
    \centering
    \includegraphics[width=\linewidth]{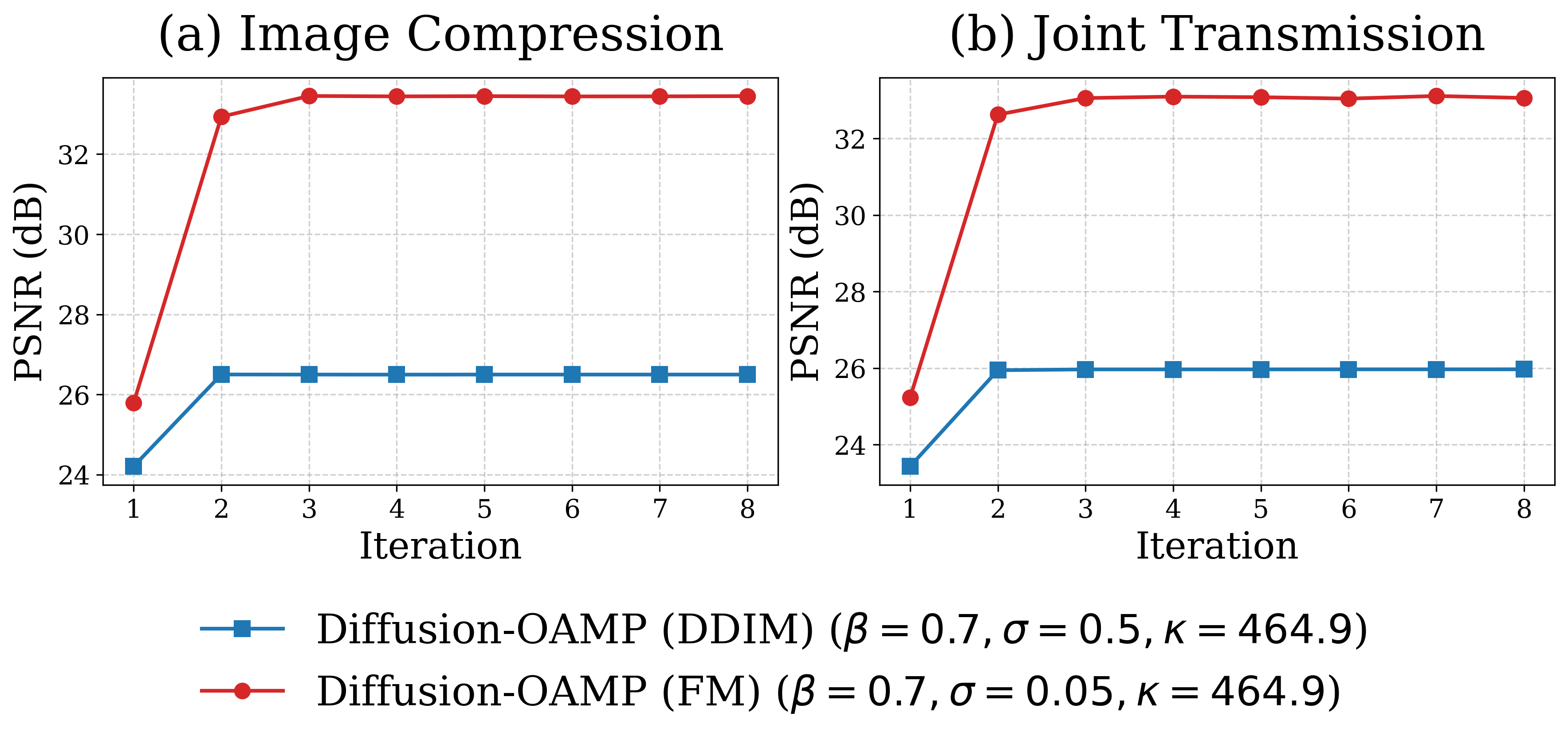}
    \caption{PSNR convergence across OAMP iterations under (a) pure AWGN and (b) TDL-A channels. Both Diffusion-OAMP variants stabilize within 3 iterations. Final converged PSNRs (DDIM / FM) are 33.15 dB / 28.52 dB in (a), and 29.56 dB / 24.36 dB in (b).}
    \label{fig:convergence}
\end{figure}

Figure~\ref{fig:convergence} evaluates the PSNR convergence behavior against the number of outer OAMP iterations under both pure image compression and joint transmission scenarios. The results show that both Diffusion-OAMP (DDIM) and Diffusion-OAMP (Flow Matching) demonstrate a rapid initial increase in PSNR, achieving stable convergence within 3 iterations across all evaluated scenarios. This fast convergence behavior verifies the operational efficiency of the proposed framework for iterative image reconstruction.

\section{Conclusion}
In this paper, we proposed Diffusion-OAMP, a training-free receiver framework for joint image compression and wireless transmission over practical fading channels. By integrating pre-trained diffusion models as non-linear estimators to regularize the severely ill-posed inverse problem, the proposed framework significantly outperforms the evaluated baselines. Specifically, it achieves higher image recovery accuracy and maintains stable structural fidelity even under highly ill-conditioned scenarios, such as low compression ratios and high noise over time-selective fading channels, while exhibiting rapid algorithmic convergence.
\bibliographystyle{IEEEtran}
\bibliography{references}
\end{document}